\documentclass[letterpaper]{article} 
\usepackage{aaai25}  
\usepackage{float}
\usepackage{newfloat}
\usepackage{multirow}
\usepackage{makecell}
\usepackage{xcolor}         
\usepackage{colortbl}
\usepackage{dsfont}
\usepackage{amsmath}
\usepackage{bbding}
\usepackage{enumitem}
\usepackage{booktabs}

\usepackage{times}  
\usepackage{helvet}  
\usepackage{courier}  
\usepackage[hyphens]{url}  
\usepackage{graphicx} 
\usepackage{natbib}  
\usepackage{caption} 
\usepackage{newfloat}
\usepackage{listings}
\DeclareCaptionStyle{ruled}{labelfont=normalfont,labelsep=colon,strut=off} 
\floatstyle{ruled}
\newfloat{listing}{tb}{lst}{}
\floatname{listing}{Listing}
\title{Preliminary Evaluation of the Test-Time Training Layers in Recommendation System (Student Abstract) }

\def\correspond{%
    \ifnum\value{eqfn}=0%
    \footnote{Corresponding authors.}%
    \setcounter{eqfn}{\value{footnote}}%
    \else%
    \footnotemark[\value{eqfn}]%
  \fi%
}%
\author{
    Tianyu Zhan,
    Zheqi Lv,
    Shengyu Zhang\correspond,
    Jiwei Li\correspond
}

\affiliations{
    Zhejiang University\\
    \{yuzt,zheqilv,sy\_zhang,jiwei\_li\}@zju.edu.cn
}

\usepackage{bibentry}

\begin{document}
\maketitle

\begin{abstract}
This paper explores the application and effectiveness of Test-Time Training (TTT) layers in improving the performance of recommendation systems. We developed a model, TTT4Rec, utilizing TTT-Linear as the feature extraction layer. Our tests across multiple datasets indicate that TTT4Rec, as a base model, performs comparably or even surpasses other baseline models in similar environments. 
\end{abstract}

%


\section{Introduction}

The rapid advancement of deep learning has significantly influenced the development of recommendation systems, leading to the emergence of several prominent sequential recommendation models, including DIN\cite{ref:din}, GRU4Rec\cite{ref:gru4rec}, SASRec\cite{ref:sasrec}, Bert4Rec\cite{sun2019bert4rec}, ComiRec\cite{ref:comirec}, ICLRec\cite{chen2022intent}, and so on.
With the advancement of technology and actual demand, recommendation technology has begun to develop toward personalization\cite{lv2023duet,lv2024semantic,lv2024intelligent,sun2022response}, multimodality\cite{zhang2021mining,zhang2022latent,zhang2023mining,ji2023online}, privacy protection\cite{liao2023ppgencdr}, cold-start\cite{liu2023joint,chen2021improving}, session-based\cite{su2023enhancing}, disentangle\cite{chen2021deep}, LLM-based\cite{zhang2023collm,bao2023tallrec}, generative-based\cite{zhao2024denoising,lin2024efficient}, and so on. 

Recently, a novel architecture known as Test-Time-Training (TTT) layers\cite{ref:ttt} has been proposed. This architecture employs self-supervised learning to update the weights of hidden states by performing gradient descent for each token. Two variants of TTT have been introduced: TTT-Linear and TTT-MLP, with TTT-Linear demonstrating superior performance and efficiency in short-context scenarios. Given the short-context nature and high-efficiency demands of recommendation systems, we conducted a preliminary exploration of TTT-Linear's performance within this context.

In this preliminary exploration, we developed a simple but effective sequential recommendation model, TTT4Rec, utilizing TTT-Linear as the feature extraction layer.

\section{Preliminaries }
TTT's central idea leverages self-supervised learning to compress historical context \(x_1, \ldots, x_t\) into a hidden state \(s_t\). Here, the context is treated as an unlabeled dataset and the state as a model. The hidden state \(s_t\) corresponds to \(W_t\), the weights of a model \(f\), which can be a linear model, a small neural network, or another form. The output is defined as:
\[ z_t = f(x_t; W_t) \]

The output token represents the prediction on \(x_t\) made by \(f\) with the updated weights \(W_t\). The update rule is a step of gradient descent on a self-supervised loss \(\ell\):
\[ W_t = W_{t-1} - \eta \nabla \ell(W_{t-1}; x_t) \]
where \(\eta\) is the learning rate.

The algorithm maps an input sequence \(x_1, \ldots, x_T\) to an output sequence \(z_1, \ldots, z_T\) using the hidden state, update rule, and output rule described above. Even during testing, the new layer trains a unique sequence of weights \(W_1, \ldots, W_T\) for each input sequence.
 \section{Model Architecture}
\begin{figure}[!h]
    \centering
    \includegraphics[width=0.86\linewidth]{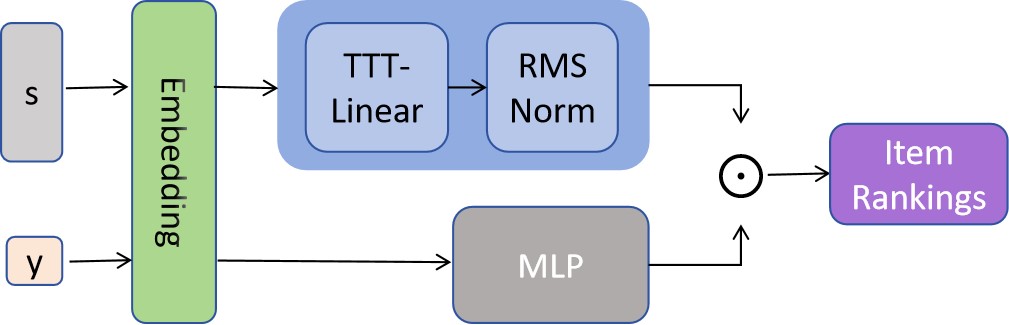}
    \caption{Model Architecture of TTT-Linear in Recommendation System}
    \label{fig:TTT4Rec}
\end{figure}
We define a data instance as \(\mathcal{X} = \{u, v, s\}\) and the corresponding label as \(\mathcal{Y} = \{y\}\), where \(u\) represents the user ID, \(v\) denotes the item ID, and \(s\) indicates the user's click sequence.Figure\ref{fig:TTT4Rec} illustrates the architecture of TTT4Rec. .
\newline
\textbf{Embedding Layers:}
Consider the user's click sequence \(s = \{v_i\}_{i=1}^{\mathcal{N}}\), with "0" serving as a padding token. For each item ID \(v_i\), latent representations are obtained through an embedding layer. This layer processes \(v_i\) to produce its ID embedding \(e_v \in \mathcal{R}^K\).
\newline
\textbf{Feature Extraction Layer:}
The set of embeddings \( E_s = \{e_1, e_2, e_3, \ldots, e_N\}_{i=1}^{\mathcal{N}} \) is then processed by the TTT-Linear layer for feature extraction. The extracted features are subsequently normalized using RMSNorm, and the hidden state \( F_s \) corresponding to the last valid click is output.
\newline
\textbf{Target Prediction:}
To conserve resources, the predicted target \(y\) is passed through the same embedding layer to retrieve the target item embedding \(e_y\). This embedding is then fed into a two-layer MLP to  extract the target features \(F_t\). Finally, the dot product of \(F_t\) with the click sequence features \(F_s\) yields the predicted click probability for the target.
\[ R = F_s \cdot F_t \]
\section{Experiment Setup}
\textbf{Datasets:} Our evaluation was carried out on three widely recognized datasets frequently used in recommendation research: Amazon Beauty, Amazon Electronics, and MovieLens-1M. For negative sampling, we adopted a 1:4 ratio during the training phase, where each positive instance is paired with four negative samples. In contrast, a 1:99 ratio was applied during the testing phase to closely approximate real-world recommendation scenarios, enhancing the robustness of our model's performance assessment.
\newline
\textbf{Baselines:} To evaluate TTT4Rec, we implemented several sequential recommendation models, including DIN\cite{ref:din}, GRU4Rec\cite{ref:gru4rec}, SASRec\cite{ref:sasrec}, and ComiRec\cite{ref:comirec}.
\newline
During the experiments, all models except ComiRec converged within the first 10 epochs. To fairly assess the performance, the experiments were conducted with a learning rate of 0.001 and 10 training epochs(50 epochs for ComiRec).

\section{Experimental Results}

\begin{table}[ht]
    \centering
    \renewcommand{\arraystretch}{1.0}
    \setlength{\tabcolsep}{1mm}
    \resizebox{0.45\textwidth}{!}{
    \begin{tabular}{c|c|c|c|c|c}
    \toprule[2pt]
    \multirow{2}{*}{\textbf{Dataset}} & \multirow{2}{*}{\textbf{Model}} & \multicolumn{4}{c}{\textbf{Metrics}} \\ \cline{3-6}
     &  & {NDCG@5}  & {NDCG@10}   & {HR@5}  & {HR@10}     \\
    \midrule
    \midrule
    \multirow{6}{*}{\texttt{Beauty}} 
    & {DIN} 
    &0.2466 & 0.2684  & 0.3546 & 0.4223 \\
    & {GRU4Rec} 
    & 0.2390&0.2692&0.3645&0.4571
 \\
    & {SASRec} 
    & 0.2512 &0.2849&0.3854&0.4893
\\
    & ComiSA
    & \textbf{0.2594} & {0.2783}  & {0.3842} & {0.4813} 
    \\
    & {ComiDR} 
    & 0.2582
    & \textbf{0.2925}
    & {0.3833}	
    & \textbf{0.4902}
    \\
    & \cellcolor[HTML]{DCDCDC}TTT4Rec
    & \cellcolor[HTML]{DCDCDC}0.2571
    &\cellcolor[HTML]{DCDCDC}0.2890
    &\cellcolor[HTML]{DCDCDC}\textbf{0.3866}
    &\cellcolor[HTML]{DCDCDC}0.4848
     
 \\
       \midrule
       \midrule
    \multirow{6}{*}{\texttt{Electronic}} & {DIN} 
    & 0.3286 &0.3664  &0.4434  &0.5603 
\\
    & {GRU4Rec} 
    & 0.3450 &0.3844 &0.4666&0.5882
 \\
    & {SASRec} 
    & 0.3420&0.3821 &0.4641&0.5881
 \\
    & ComiSA 
    & {0.3065}	& {0.3301}	& {0.4221}	& {0.5418}	
    \\
    & ComiDR 
    & {0.3228}	
    & {0.3623}	
    & {0.4414}	
    & {0.5637}	
\\
    & \cellcolor[HTML]{DCDCDC}{TTT4Rec}
    &\cellcolor[HTML]{DCDCDC}\textbf{0.3486}	& \textbf{\cellcolor[HTML]{DCDCDC}0.3892}		& \cellcolor[HTML]
    {DCDCDC}\textbf{0.4712}	& \cellcolor[HTML]
    {DCDCDC}\textbf{0.5968}		 \\
       \midrule
       \midrule

    \multirow{6}{*}{\texttt{ML-1m}} & {DIN} 
    & 
    0.4703 &0.5081 &0.6318&0.7485
 \\
    & {GRU4Rec} 
    & 
    0.5209 &0.5566 &0.6762 &0.7863 
 \\
    & {SASRec} 
    & 
    0.5217 & 0.5574 & 0.6790 & \textbf{0.7919} 
\\
    & ComiSA 
    & {0.4040}	& {0.4450}	& {0.5417}	& {0.6982}		 \\

    & {ComiDR} 
    & {0.4072}	& {0.4549}	& {0.5550}	& {0.7022}	 \\

    &\cellcolor[HTML]{DCDCDC} TTT4Rec
    &\cellcolor[HTML]{DCDCDC}\textbf{0.5239}	
    &\cellcolor[HTML]{DCDCDC}\textbf{0.5577}	
    &\cellcolor[HTML]{DCDCDC}\textbf{0.6834}	
    &\cellcolor[HTML]{DCDCDC}0.7877
     \\
    \bottomrule[2pt]
    \end{tabular}
    }
\caption{ Performance comparison of different methods. The best performance is highlighted in bold.}
\label{table:main_table}
\end{table}

\begin{figure}[t]
    \centering
    \includegraphics[width=1\linewidth]{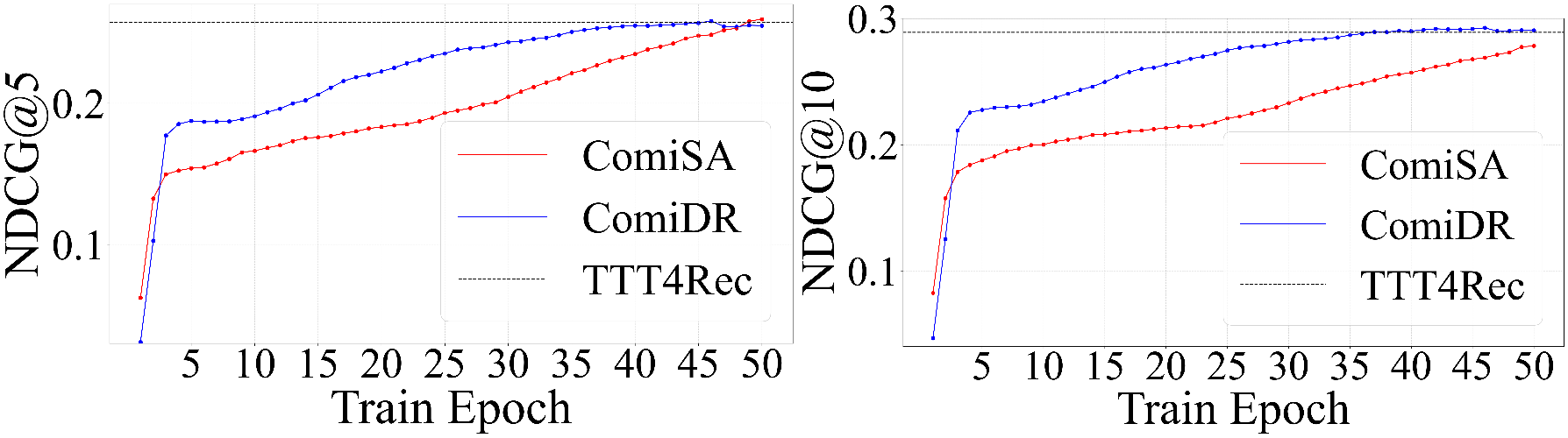}
    \caption{The NDCG@5 metrics for ComiSA and ComiDR evolve during training, with the dashed line showing TTT4Rec's performance at epoch 1 on the Beauty dataset.  }
    \label{fig:ndcg}
\end{figure}

\begin{figure}[!h]
    \centering
    \includegraphics[width=1\linewidth]{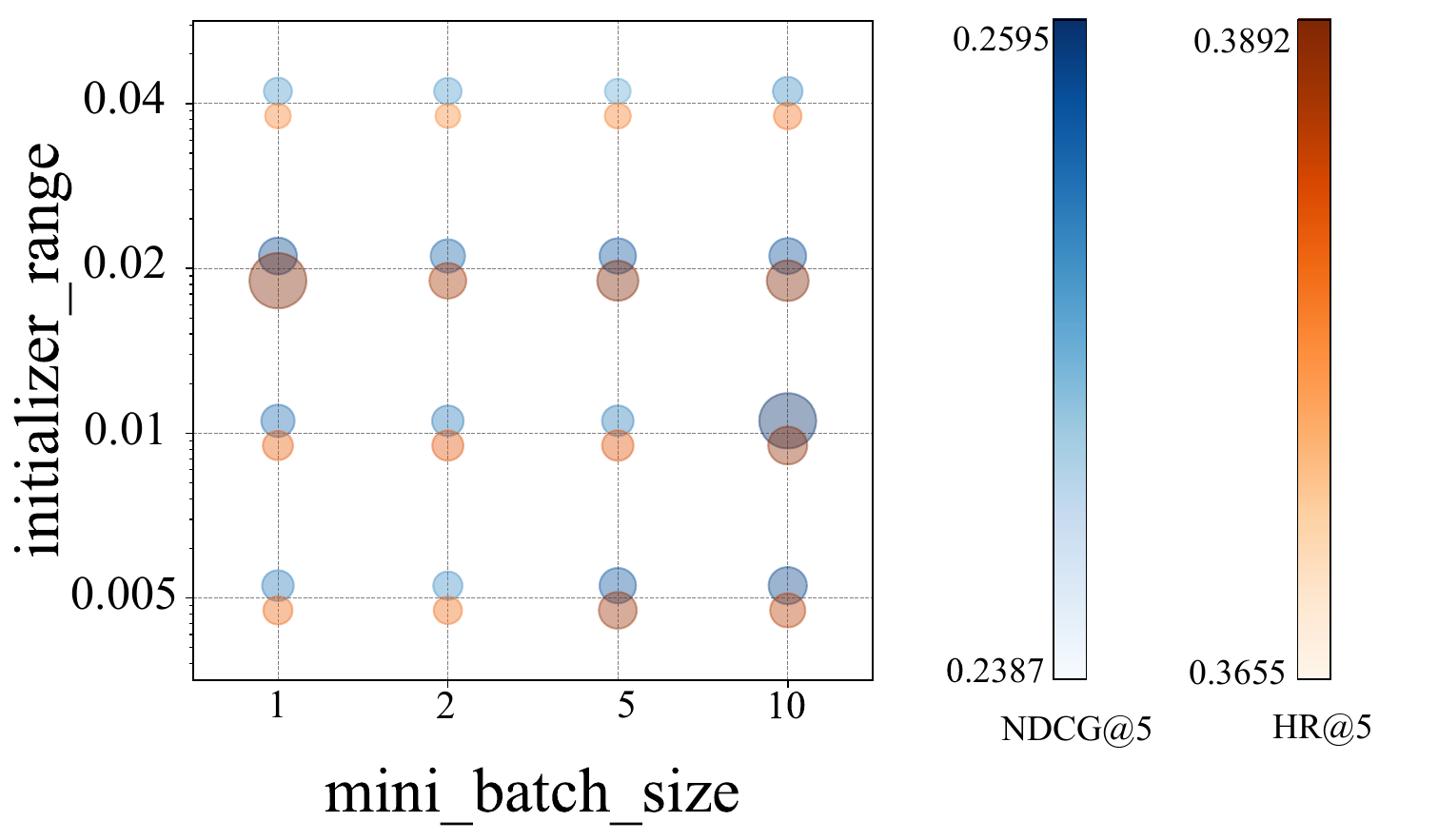}
    \caption{Hyperparameter Grid Search.(Beauty Dataset) }
    \label{fig:hyper}
\end{figure}

\textbf{Overall Performance:} Table \ref{table:main_table} summarizes the evaluations of TTT4Rec and baseline models across the datasets. TTT4Rec matches or surpasses the baselines in NDCG and Hit metrics. While ComiRec performs better overall due to advanced techniques, its training efficiency is lower, requiring 50 epochs on the Beauty dataset to match the performance TTT4Rec achieves in a single epoch (Figure \ref{fig:ndcg}). 
\\
\textbf{Hyper-parameter Analysis:} In our experiments, we identified two parameters that had a significant impact on performance, as shown in Figure \ref{fig:hyper}:

\begin{enumerate}
    \item \textbf{initializer\_range:} This parameter sets the standard deviation for initializing TTT layer parameters. A value that's too small leads to minimal parameter variation, impairing functionality, while a value that's too large makes the parameter space too broad, complicating the search for optimal parameters. 
    \item \textbf{mini\_batch\_size:}This parameter represents the number of tokens processed simultaneously in the mini-batch TTT parallelization strategy. The results suggest that optimal performance is achieved with batch sizes of 1 or 10, indicating that both online and batch gradient descent (GD) are more effective for TTT-Linear in this context. Mini-batch GD updates the current token using W derived from several tokens apart, which may link unrelated tokens and adversely affect performance. 
\end{enumerate}

\section{Conclusion and Future Work}
As a base model utilizing TTT-Linear for feature extraction, TTT4Rec has shown performance that is on par with, or even exceeds, other base models within the same environment. Future research will aim to further refine the parameter W in TTT-Linear. 

\newpage
\section{Acknowledgements}
This work was in part supported by 2030 National Science and Technology Major Project (2022ZD0119100), National Natural Science Foundation of China (No. 62402429, 62441605), the Key Research and Development Program of Zhejiang Province (No. 2024C03270), ZJU Kunpeng\&Ascend Center of Excellence.

\bibliography{aaai25}

\begin{thebibliography}{24}
\providecommand{\natexlab}[1]{#1}

\bibitem[{Bao et~al.(2023)Bao, Zhang, Zhang, Wang, Feng, and He}]{bao2023tallrec}
Bao, K.; Zhang, J.; Zhang, Y.; Wang, W.; Feng, F.; and He, X. 2023.
\newblock Tallrec: An effective and efficient tuning framework to align large language model with recommendation.
\newblock In \emph{Proceedings of the 17th ACM Conference on Recommender Systems}, 1007--1014.

\bibitem[{Cen et~al.(2020)Cen, Zhang, Zou, Zhou, Yang, and Tang}]{ref:comirec}
Cen, Y.; Zhang, J.; Zou, X.; Zhou, C.; Yang, H.; and Tang, J. 2020.
\newblock Controllable multi-interest framework for recommendation.
\newblock In \emph{Proceedings of the 26th ACM SIGKDD International Conference on Knowledge Discovery \& Data Mining}, 2942--2951.

\bibitem[{Chen et~al.(2022)Chen, Liu, Li, McAuley, and Xiong}]{chen2022intent}
Chen, Y.; Liu, Z.; Li, J.; McAuley, J.; and Xiong, C. 2022.
\newblock Intent contrastive learning for sequential recommendation.
\newblock In \emph{Proceedings of the ACM Web Conference 2022}, 2172--2182.

\bibitem[{Chen, Wang, and Yin(2021)}]{chen2021improving}
Chen, Z.; Wang, D.; and Yin, S. 2021.
\newblock Improving cold-start recommendation via multi-prior meta-learning.
\newblock In \emph{Advances in Information Retrieval: 43rd European Conference on IR Research, ECIR 2021, Virtual Event, March 28--April 1, 2021, Proceedings, Part II 43}, 249--256. Springer.

\bibitem[{Chen, Xu, and Wang(2021)}]{chen2021deep}
Chen, Z.; Xu, Z.; and Wang, D. 2021.
\newblock Deep transfer tensor decomposition with orthogonal constraint for recommender systems.
\newblock In \emph{Proceedings of the AAAI Conference on Artificial Intelligence}, volume~35, 4010--4018.

\bibitem[{Hidasi et~al.(2016)Hidasi, Karatzoglou, Baltrunas, and Tikk}]{ref:gru4rec}
Hidasi, B.; Karatzoglou, A.; Baltrunas, L.; and Tikk, D. 2016.
\newblock Session-based recommendations with recurrent neural networks.
\newblock \emph{International Conference on Learning Representations 2016}.

\bibitem[{Ji et~al.(2023)Ji, Liu, Zhang, Wei, Ni, and Wang}]{ji2023online}
Ji, W.; Liu, X.; Zhang, A.; Wei, Y.; Ni, Y.; and Wang, X. 2023.
\newblock Online distillation-enhanced multi-modal transformer for sequential recommendation.
\newblock In \emph{Proceedings of the 31st ACM International Conference on Multimedia}, 955--965.

\bibitem[{Kang and McAuley(2018)}]{ref:sasrec}
Kang, W.-C.; and McAuley, J. 2018.
\newblock Self-attentive sequential recommendation.
\newblock In \emph{2018 IEEE International Conference on Data Mining (ICDM)}, 197--206. IEEE.

\bibitem[{Liao et~al.(2023)Liao, Liu, Zheng, Yao, and Chen}]{liao2023ppgencdr}
Liao, X.; Liu, W.; Zheng, X.; Yao, B.; and Chen, C. 2023.
\newblock Ppgencdr: A stable and robust framework for privacy-preserving cross-domain recommendation.
\newblock In \emph{Proceedings of the AAAI Conference on Artificial Intelligence}, volume~37, 4453--4461.

\bibitem[{Lin et~al.(2024)Lin, Yang, Wang, Li, Du, Feng, Ng, and Chua}]{lin2024efficient}
Lin, X.; Yang, C.; Wang, W.; Li, Y.; Du, C.; Feng, F.; Ng, S.-K.; and Chua, T.-S. 2024.
\newblock Efficient Inference for Large Language Model-based Generative Recommendation.

\bibitem[{Liu et~al.(2023)Liu, Zheng, Chen, Su, Liao, Hu, and Tan}]{liu2023joint}
Liu, W.; Zheng, X.; Chen, C.; Su, J.; Liao, X.; Hu, M.; and Tan, Y. 2023.
\newblock Joint internal multi-interest exploration and external domain alignment for cross domain sequential recommendation.
\newblock In \emph{Proceedings of the ACM Web Conference 2023}, 383--394.

\bibitem[{Lv et~al.(2024{\natexlab{a}})Lv, He, Zhan, Zhang, Zhang, Chen, Zhao, and Wu}]{lv2024semantic}
Lv, Z.; He, S.; Zhan, T.; Zhang, S.; Zhang, W.; Chen, J.; Zhao, Z.; and Wu, F. 2024{\natexlab{a}}.
\newblock Semantic Codebook Learning for Dynamic Recommendation Models.
\newblock In \emph{Proceedings of the 32nd ACM International Conference on Multimedia}, 9611--9620.

\bibitem[{Lv et~al.(2024{\natexlab{b}})Lv, Zhang, Chen, Zhang, and Kuang}]{lv2024intelligent}
Lv, Z.; Zhang, W.; Chen, Z.; Zhang, S.; and Kuang, K. 2024{\natexlab{b}}.
\newblock Intelligent model update strategy for sequential recommendation.
\newblock In \emph{Proceedings of the ACM on Web Conference 2024}, 3117--3128.

\bibitem[{Lv et~al.(2023)Lv, Zhang, Zhang, Kuang, Wang, Wang, Chen, Shen, Yang, Ooi et~al.}]{lv2023duet}
Lv, Z.; Zhang, W.; Zhang, S.; Kuang, K.; Wang, F.; Wang, Y.; Chen, Z.; Shen, T.; Yang, H.; Ooi, B.~C.; et~al. 2023.
\newblock Duet: A tuning-free device-cloud collaborative parameters generation framework for efficient device model generalization.
\newblock In \emph{Proceedings of the ACM Web Conference 2023}, 3077--3085.

\bibitem[{Su et~al.(2023)Su, Chen, Liu, Wu, Zheng, and Lyu}]{su2023enhancing}
Su, J.; Chen, C.; Liu, W.; Wu, F.; Zheng, X.; and Lyu, H. 2023.
\newblock Enhancing hierarchy-aware graph networks with deep dual clustering for session-based recommendation.
\newblock In \emph{Proceedings of the ACM Web Conference 2023}, 165--176.

\bibitem[{Sun et~al.(2019)Sun, Liu, Wu, Pei, Lin, Ou, and Jiang}]{sun2019bert4rec}
Sun, F.; Liu, J.; Wu, J.; Pei, C.; Lin, X.; Ou, W.; and Jiang, P. 2019.
\newblock BERT4Rec: Sequential recommendation with bidirectional encoder representations from transformer.
\newblock In \emph{Proceedings of the 28th ACM international conference on information and knowledge management}, 1441--1450.

\bibitem[{Sun et~al.(2022)Sun, Wang, Song, Feng, and Nie}]{sun2022response}
Sun, T.; Wang, C.; Song, X.; Feng, F.; and Nie, L. 2022.
\newblock Response generation by jointly modeling personalized linguistic styles and emotions.
\newblock \emph{ACM Transactions on Multimedia Computing, Communications, and Applications (TOMM)}, 18(2): 1--20.

\bibitem[{Sun et~al.(2024)Sun, Li, Dalal, Xu, Vikram, Zhang, Dubois, Chen, Wang, Koyejo, Hashimoto, and Guestrin}]{ref:ttt}
Sun, Y.; Li, X.; Dalal, K.; Xu, J.; Vikram, A.; Zhang, G.; Dubois, Y.; Chen, X.; Wang, X.; Koyejo, S.; Hashimoto, T.; and Guestrin, C. 2024.
\newblock Learning to (Learn at Test Time): RNNs with Expressive Hidden States.
\newblock arXiv:2407.04620.

\bibitem[{Zhang et~al.(2023{\natexlab{a}})Zhang, Liu, Wu, and Wang}]{zhang2023mining}
Zhang, J.; Liu, Q.; Wu, S.; and Wang, L. 2023{\natexlab{a}}.
\newblock Mining Stable Preferences: Adaptive Modality Decorrelation for Multimedia Recommendation.
\newblock In \emph{Proceedings of the 46th International ACM SIGIR Conference on Research and Development in Information Retrieval}, 443--452.

\bibitem[{Zhang et~al.(2021)Zhang, Zhu, Liu, Wu, Wang, and Wang}]{zhang2021mining}
Zhang, J.; Zhu, Y.; Liu, Q.; Wu, S.; Wang, S.; and Wang, L. 2021.
\newblock Mining latent structures for multimedia recommendation.
\newblock In \emph{Proceedings of the 29th ACM international conference on multimedia}, 3872--3880.

\bibitem[{Zhang et~al.(2022)Zhang, Zhu, Liu, Zhang, Wu, and Wang}]{zhang2022latent}
Zhang, J.; Zhu, Y.; Liu, Q.; Zhang, M.; Wu, S.; and Wang, L. 2022.
\newblock Latent structure mining with contrastive modality fusion for multimedia recommendation.
\newblock \emph{IEEE Transactions on Knowledge and Data Engineering}, 35(9): 9154--9167.

\bibitem[{Zhang et~al.(2023{\natexlab{b}})Zhang, Feng, Zhang, Bao, Wang, and He}]{zhang2023collm}
Zhang, Y.; Feng, F.; Zhang, J.; Bao, K.; Wang, Q.; and He, X. 2023{\natexlab{b}}.
\newblock Collm: Integrating collaborative embeddings into large language models for recommendation.
\newblock \emph{arXiv preprint arXiv:2310.19488}.

\bibitem[{Zhao et~al.(2024)Zhao, Wenjie, Xu, Sun, Feng, and Chua}]{zhao2024denoising}
Zhao, J.; Wenjie, W.; Xu, Y.; Sun, T.; Feng, F.; and Chua, T.-S. 2024.
\newblock Denoising diffusion recommender model.
\newblock In \emph{Proceedings of the 47th International ACM SIGIR Conference on Research and Development in Information Retrieval}, 1370--1379.

\bibitem[{Zhou et~al.(2018)Zhou, Zhu, Song, Fan, Zhu, Ma, Yan, Jin, Li, and Gai}]{ref:din}
Zhou, G.; Zhu, X.; Song, C.; Fan, Y.; Zhu, H.; Ma, X.; Yan, Y.; Jin, J.; Li, H.; and Gai, K. 2018.
\newblock Deep interest network for click-through rate prediction.
\newblock In \emph{Proceedings of the 24th ACM SIGKDD International Conference on Knowledge Discovery \& Data Mining}, 1059--1068.

\end{thebibliography}

\end{document}